\documentclass[
twocolumn, preprintnumbers, nofootinbib,amsmath,amssymb]{revtex4}

\usepackage{graphicx}
\usepackage{dcolumn}
\usepackage{bm}

\newcommand\ee{\end{equation}}
\newcommand\be{\begin{equation}}
\newcommand\eea{\end{eqnarray}}
\newcommand\bea{\begin{eqnarray}}

\newcommand\mpl{M_{\rm Pl}}

\begin{document}


\title{Conformal invariance of scalar perturbations in inflation}

\author{Paolo Creminelli}
\email{creminel@ictp.it}

\affiliation{%
Abdus Salam International Centre for Theoretical Physics\\ Strada Costiera 11, 34151, Trieste, Italy
}%

\date{\today}

\begin{abstract}
In inflationary models where the source of scalar perturbations is not the inflaton, but one or more scalars with negligible coupling with the inflaton, the resulting perturbations are not only scale invariant, but fully conformally invariant with conformal dimension close to zero. This is closely related to the fact that correlation functions can only depend on the de Sitter invariant distances. These properties follow from the isometries of the inflationary de Sitter space and are thus completely independent of the dynamics. The 3-point function is fixed in terms of two constants, while the 4-point function is a function of two parameters (instead of five as in the absence of conformal invariance). The conformal invariance of correlators can be directly checked in Fourier space, as we show in an explicit example. A detection of a non-conformal correlation function, for example an equilateral 3-point function, would imply that the source of perturbations is not decoupled from the inflaton.
\end{abstract}

\maketitle


\noindent

Inflation takes place in (an approximate) de Sitter space
\be
ds^2 = \frac1{H^2\eta^2} (- d \eta^2 + d \vec x^2) \;,
\ee
whose isometry group is $SO(4,1)$. The time-evolving inflaton background is homogeneous and rotationally invariant, so that translations and rotations are good symmetries of the whole system. The dilation isometry $\eta \to \lambda \eta$, $\vec x \to \lambda \vec x$ is also an approximate symmetry of the inflaton background in the limit in which its dynamics varies slowly in time. It is this isometry which guarantees a scale invariant spectrum, independently of the inflaton dynamics. In Fourier space dilations act as 
\be
\varphi_{\vec k} \to \lambda^{-3} \varphi_{\vec k/\lambda} \;,
\ee 
so that the 2-point function is constrained to be of the form
\be
\langle \varphi_{\vec k_1} \varphi_{\vec k_2} \rangle = (2\pi)^3 \delta(\vec k_1 + \vec k_2) \frac{1}{k_1^3} F(k_1 \eta) \;.
\ee
If perturbations become time-independent when out of the Hubble radius, the function $F$ must be a constant in this limit and this gives a scale invariant spectrum\footnote{This argument for scale invariance appeared in \cite{Creminelli:2010ba}.}. 

Besides translations, rotations and dilations, de Sitter possesses three additional isometries 
\be
\label{eq:special}
\eta \to \eta - 2 \eta (\vec b \cdot \vec x) \;,\quad x^i \to x^i + b^i(-\eta^2 + \vec x^2) -2 x^i (\vec b \cdot \vec x) \;,
\ee
parametrized by the infinitesimal vector $\vec b$.
These isometries, however, are not a symmetry of the inflaton background as surfaces of constant inflaton value, i.e.~of constant $\eta$, are not mapped into themselves.  Therefore inflationary correlation functions will not, in general, possess any additional symmetry.

In this paper we point out that, when the leading mechanism which generates primordial perturbations is not the inflaton, but a different sector with negligible coupling with the inflaton, then the de Sitter isometries imply that the scalar correlation functions are not only scale invariant, but they enjoy full conformal symmetry, with an approximately zero conformal weight. This paper build on \cite{Maldacena:2011nz} where the de Sitter isometries have been used to constrain the general form of the 3-point function of gravitational waves and on \cite{Antoniadis:2011ib}, which discusses the conformal symmetry of scalar perturbations. Here we give physical examples in which this symmetry is realized and we will see that scalar perturbations have conformal weight (close to) zero. Notice that our arguments are simply the de Sitter analogue of the standard statement that correlation function in Minkowski are Poincar\'e invariant.

The scenario that we want to study is quite general. During inflation we have a sector with negligible interactions with the inflaton\footnote{This separation is technically natural: the only mandatory interaction between the two sectors is gravity, so that loop effects will be small, as suppressed by powers of $\mpl$.}, composed of particles of arbitrary spin, mass and interactions. We are interested in the late time correlation functions of scalars, in particular in the ones with a mass much smaller than $H$, as the others will decay quickly when out of the Hubble radius. When we are far out of the Hubble radius the transformation of the spatial coordinates in the isometry \eqref{eq:special} becomes
\be
x^i \to x^i + b^i \vec x^2 -2 x^i (\vec b \cdot \vec x)
\ee
which is a special conformal transformation of the spatial coordinates. In this regime a scalar $\varphi$ with mass $m$ evolves in $\eta$ as 
\be
\label{eq:delta}
\varphi \sim \eta^{\Delta} \;, \quad  \Delta = \frac32 \left(1-\sqrt{1- \frac{4 m^2}{9H^2}}\right) \;. 
\ee
Thus the $\eta$ transformation in eq.~\eqref{eq:special} acts the same way a conformal transformation would act on a primary field of dimension $\Delta$. The same holds for dilations and thus we conclude that the late time correlation functions, at equal time, are conformal invariant with conformal weight $\Delta$. For fields with small mass we have $\Delta \ll 1$ (\footnote{Notice that one cannot consistently keep the deviations from scale-invariance induced by the small mass, as the corrections due to the deviation from exact de Sitter are in general comparable and cannot be accounted for in this context.}).
Notice that the logic above is only based on symmetry and therefore does not even require that the sector is weakly coupled at the scale $H$. This is completely analogous to what discussed in \cite{Maldacena:2011nz} for tensor modes.

An equivalent way to see these additional constraints is by noticing that correlation functions in de Sitter must be functions of the de Sitter invariant distance (see for example \cite{Spradlin:2001pw}), which for a pair of points $(\eta_i, \vec x_i)$ and $(\eta_j, \vec x_j)$ reads
\be
\frac{|\vec x_i -\vec x_j|^2}{\eta_i \eta_j} - \left(\frac{\eta_i}{\eta_j} + \frac{\eta_j}{\eta_i}\right) \;.
\ee
In the limit in which points are at a distance much larger than the Hubble radius, the second term can be neglected so that correlation functions must depend on $|\vec x_i -\vec x_j|^2/(\eta_i \eta_j)$. This implies that the time-dependence of the fields in the long wavelength regime -- which is related to their mass -- also fixes the dependence on the spatial coordinates.

Perturbations of these scalars can easily feed into adiabatic perturbations, through one of the standard conversion mechanisms (curvaton \cite{hep-ph/0109214,Lyth:2001nq}, inhomogeneous reheating \cite{Dvali:2003em}...): using the $\delta N$ formalism \cite{Lyth:2004gb}, the adiabatic curvature $\zeta$ will be a local function of these scalar perturbations. This introduces additional non-Gaussianities in the observable correlation functions of $\zeta$. However we will see that also these contributions are conformally invariant, so that also the resulting correlation functions of $\zeta$ are conformally invariant. 

As the 2-point function is already fixed by dilation symmetry, the constraints of conformal invariance are only useful for the higher-order correlators. In the following we study the 3-point and the 4-point functions.

{\bf 3-point function}. Let us consider the 3-point function induced by an interaction
\be
\label{eq:cubic}
- \int d^4 x \sqrt{-g} \;\frac{M}{6} \varphi^3 \;.
\ee
Eventually we will be interested in the case of a massless scalar, but let us start assuming that $\varphi$ has a mass $m = \sqrt{2} H$, so that $\Delta =1$ in eq.~\eqref{eq:delta}, as the algebra is particularly simple in this case. Using the standard in-in formalism we get the 3-point function of $\varphi$
\be
\langle\varphi_{\vec k_1}\varphi_{\vec k_2}\varphi_{\vec k_3} \rangle = (2\pi)^3 \delta(\sum_i \vec k_i) \frac{\pi}{8} M H^2 \eta_*^3 \cdot \frac{1}{k_1 k_2 k_3} 
\ee
where $\eta_*$ corresponds to the time when the 3-point function is evaluated. Going to real space the 3-point function reads\footnote{Here and in the following we used the results of \cite{Antoniadis:2011ib} to move back and forth from Fourier space.}
\be
\label{eq:delta1real}
\langle\varphi (\vec x_1)\varphi(\vec x_2)\varphi(\vec x_3) \rangle = \frac{M H^2 \eta_*^3}{64 \pi^2} \cdot \frac{1}{|\vec x_1 - \vec x_2||\vec x_1 - \vec x_3||\vec x_2 - \vec x_3|} \;.
\ee
We see that indeed this corresponds to the 3-point function of a scalar with $\Delta =1$ in a 3d conformal field theory. The same result can be obtained using that the 3-point function can only depend on the de Sitter invariant distance $|\vec x_i -\vec x_j|^2/(\eta_i \eta_j)$. Notice, first of all, that it is not possible with just three points to build from this a time-independent quantity (here we are not restricting to the same $\eta$ for the three points). As the long-wavelength time dependence is fixed by the mass of the field to be $\propto \eta_1 \eta_2 \eta_3$, this fixes the spatial dependence to be proportional to eq.~\eqref{eq:delta1real}. For general $\Delta$, the 3-point function will behave as 
\be
\label{eq:Delta}
\langle\varphi (\vec x_1)\varphi(\vec x_2)\varphi(\vec x_3) \rangle \propto \frac{1}{|\vec x_1 - \vec x_2|^\Delta|\vec x_1 - \vec x_3|^\Delta|\vec x_2 - \vec x_3|^\Delta} \;.
\ee

Let us now look at a massless scalar. In this case the result is \cite{Zaldarriaga:2003my} \footnote{The result in this paper contains a typo, as point out in \cite{Seery:2008qj}.}
\be
\begin{split}
\label{eq:matias}
& \langle\varphi_{\vec k_1}\varphi_{\vec k_2}\varphi_{\vec k_3} \rangle =  (2\pi)^3 \delta(\sum_i \vec k_i)  \frac{H^2}{\prod_i 2 k_i^3} \frac{2 M}{3} \cdot \\ & \left[\sum_i k_i^3 (-1 + \gamma + \log(-k_t \eta_*)) + k_1 k_2 k_3 - \sum_{i \neq j} k_i^2 k_j \right]  \;,
\end{split}
\ee
where $k_t \equiv  \sum_i k_i$ and $\gamma$ is the Euler's constant. Notice that $(-1+\gamma +\log(-\eta_*))$ multiplies an exactly `local' term, i.e.~of the form $k_1^{-3}k_2^{-3}$ + perm. 
The Fourier transform of this expression reads
\be
\begin{split}
\label{eq:log3}
& \langle\varphi_1(\vec x_1)\varphi_2(\vec x_2)\varphi_3(\vec x_3) \rangle = \\ &-\frac{M H^2}{48 \pi^4}\log\frac{|\vec x_1 - \vec x_2|}{A \eta_*} \log\frac{|\vec x_1 - \vec x_3|}{A \eta_*} \log\frac{|\vec x_2 - \vec x_3|}{A \eta_*} \;,
\end{split}
\ee
where $A$ is a constant which can be evaluated numerically: $A \simeq 2.95$.  As expected the result only depends on the de Sitter invariant distances between pair of points. To go to real space, we used the results of \cite{Antoniadis:2011ib}, which studies the Fourier transform of the general conformal 3-point function \eqref{eq:Delta}. In the limit $\Delta \to 0$ each power in eq.~\eqref{eq:Delta} can be expanded as $|\vec x_i - \vec x_j|^{-\Delta} \simeq 1 - \Delta \log{|\vec x_i - \vec x_j|} $ and we get (neglecting terms which do not contribute to the Fourier transform at non-zero $k$'s) a local shape, product of two logarithms in real space, and, suppressed by $\Delta$, a product of three logarithms. In Fourier space, one can get rid of local terms by taking the derivative with respect to all the three $k$'s: in doing this the local shape, whose three terms depend only on two momenta, cancels. In this way we can check that the Fourier transform of \eqref{eq:log3}, coincides with \eqref{eq:matias}, up to a local, $\eta$-independent term, which corresponds to $A$, whose value can be numerically determined.

Notice that eq.~\eqref{eq:log3} is not simply the limit $\Delta \to 0$ of the general conformal 3-point function \eqref{eq:Delta}. The reason is that the two limits $\eta \to 0$ and $\Delta \to 0$ do not commute, as the result \eqref{eq:Delta} contains corrections of the form $(-\eta)^\Delta$, which are sensitive to the order of the limits. More physically, a massless scalar, which would be $\eta$-independent out of the Hubble radius, acquires a logarithmic time dependence because of the cubic interaction. This logarithmic time dependence, through de Sitter invariance, fixes the 3-point function to have the form \eqref{eq:log3}.

Conformal invariance does not fix the constant $A$, as a local shape, which is a product of two logarithms in real space, is separately invariant on its own.
The 3-point function is determined up to two constants: the overall normalization and the local contribution. 
It is important to stress that another source of local non-Gaussianity, which will contribute to the constant $A$, comes from the non-linear relation between the scalar (or scalars) $\varphi$ and $\zeta$. This gives a 3-point function for $\zeta$ even in the absence of sizable cubic interactions for $\varphi$.

The strong restriction imposed by conformal invariance on the 3-point function is at first puzzling. We can write an infinite set of cubic interactions involving scalars: how is possible that we always end up with the two shapes discussed above? Let us see that one can always reduce to zero the number of derivatives in a cubic interactions, using field redefinitions\footnote{A similar argument restricted to the lowest order operators appeared in \cite{Senatore:2010wk}.}. Consider for instance an interaction among three scalars of the form 
\be
\label{eq:dpi2pi}
\frac{1}{M}\int d^4 x \sqrt{-g} \nabla_\mu \varphi_1 \nabla^\mu \varphi_2 \varphi_3 \;.
\ee
Integrating by parts this can be rewritten as
\be
\label{eq:dpi2pi2}
\frac{1}{M}\int d^4 x \sqrt{-g}\frac12 (\Box\varphi_3 \varphi_1 \varphi_2 - \Box\varphi_1 \varphi_2 \varphi_3 - \Box\varphi_2 \varphi_1 \varphi_3) \;.
\ee
Now the interactions are proportional to the quadratic equation of motion (without a mass term) and therefore we can reduce the number of derivatives doing a field redefinition. For example the field redefinition $\varphi_3 \to \varphi_3 - \frac1{2 M} \varphi_1 \varphi_2$ eliminates the first interaction -- or gets rid of its derivatives in the presence of a mass term -- and analogously for the others. This field redefinition contributes a 3-point function with a local shape, so that the operator \eqref{eq:dpi2pi} does not give a new shape. This argument works also if additional derivatives are acting on the scalars. First of all notice that we do not have to worry about commuting derivatives: in de Sitter the Riemann tensor can be written in terms of the metric, so that it  just gives new operators with less derivatives to which the arguments can be recursively applied. Consider two covariant derivatives whose indices are contracted. If these derivatives act on different fields, one can do the same as we did to go from eq.~\eqref{eq:dpi2pi} to eq.~\eqref{eq:dpi2pi2}, so that they end up on the same field. Then the operator is proportional to the second order equations of motion and can be removed by a field redefinition. If the required field redefinition is local, we get the usual local shape, while if it contains derivatives it does not contribute to the asymptotic 3-point function as derivatives decay out of the Hubble radius.  Therefore any cubic operator will only contribute to the two shapes discussed above. Notice that the argument does not go through for higher correlators: the step from eq.~\eqref{eq:dpi2pi} to eq.~\eqref{eq:dpi2pi2} only works with three fields.

In eq.~\eqref{eq:matias} the logarithm will be rather large, as it describes the number of e-folds of evolution outside the Hubble radius, until $\varphi$ is finally converted to $\zeta$. This will dominate the 3-point function, which will thus be quite close to a local shape. On symmetry grounds, we can have a general linear combination of the two conformal invariant shapes. In the squeezed limit (in which one of the modes, $k_L$, is much smaller than the other two, $k_S$) this general shape goes as $k_S^{-3} k_L^{-3} \log{k_S/k_*}$, where $k_*$ is a fixed scale. In the lucky case in which $k_*$ lies in the observable window, one would have an interesting effect for the scale-dependent bias \cite{Dalal:2007cu}, as the bias would depend logarithmically on the size of the objects we are looking at, which fixes the short scale $k_S$.

{\bf 4-point function}. Let us consider for instance a massless, derivatively coupled scalar in de Sitter with a quartic Hamiltonian coupling 
\be
\int d^4x \frac{1}{8 M^4} (\partial_\mu\varphi)^2 (\partial_\nu\varphi)^2 = \int d^3x d\eta \frac1{8 M^4}(\varphi'^2-(\partial_i\varphi)^2)^2 \;.
\ee
It is straightforward to calculate the 4-point function induced by this interaction using the standard in-in formalism. The result is
\begin{multline} 
\langle\varphi_{\vec k_1}\varphi_{\vec k_2}\varphi_{\vec k_3} \varphi_{\vec k_4}\rangle = (2\pi)^3 \delta(\sum_i \vec k_i) \frac1{M^4} \frac{H^8}{\prod_i 2 k_i^3} \cdot \\ \Bigg[-\frac{144 k_1^2 k_2^2 k_3^2 k_4^2}{k_t^5} - 4 \left(\frac{12 k_1 k_2 k_3 k_4}{k_t^5} +\frac{3 \prod_{i<j<l}k_i k_j k_l}{k_t^4}+ \right. \\ \left.\frac{\prod_{i<j} k_i k_j}{k_t^3} +\frac{1}{k_t}\right)  \big((\vec k_1 \cdot \vec k_2) (\vec k_3 \cdot \vec k_4) + {\rm 2\;perm.}\big) \\ + (\vec k_1 \cdot \vec k_2) \left(\frac{4 k_3^2 k_4^2}{k_t^3}+\frac{12 (k_1+k_2) k_3^2 k_4^2}{k_t^4}+\frac{48 k_1 k_2 k_3^2
   k_4^2}{k_t^5}\right)  \\ + {\rm 5\;perm.} \Bigg]  \label{eq:dp4}\;,
\end{multline}
where the permutations act only on the last term. Of course there is little hope of Fourier transforming to real space. However the constraints of conformal invariance can be checked directly in Fourier space, as explained in \cite{Maldacena:2011nz}. Dilation invariance is trivial, while for the special conformal transformations of a $\Delta =0$ field we have to verify that 
\begin{multline}\label{eq:conf4}
\sum_{a=1,2,3,4} \left[6 \vec b \cdot \vec\partial_{k_a} - \vec b \cdot \vec k_a \vec \partial^{\;2}_{k_a} + 2 \vec k_a \cdot \vec \partial_{k_a} (\vec b \cdot \vec\partial_{k_a}) \right]  \\\langle\varphi_{\vec k_1}\varphi_{\vec k_2}\varphi_{\vec k_3} \varphi_{\vec k_4}\rangle' =0 \;,
\end{multline}
for any vector $\vec b$.  The prime indicates we have removed the Dirac delta from expression \eqref{eq:dp4} and the equality holds only when the four vectors $\vec k_a$ sum up to zero. It is straightforward, but lengthy, to verify this for \eqref{eq:dp4}. One can also check that the invariance would not hold if one considers separately the operators $\varphi'^4$, $\varphi'^2 (\partial_i\varphi)^2$ and $(\partial_i\varphi)^2 (\partial_i\varphi)^2$. These operators appear separately when considering the inflaton Lagrangian, as there is a privileged time slicing.

Additional contributions to the 4-point function will come from the non-linear relation between $\zeta$ and the scalars, set by the details of the conversion mechanism. This induces two possible shapes, conventionally proportional to $g_{\rm NL}$ and $\tau_{\rm NL}$ \cite{Byrnes:2006vq}. It is straightforward to verify that both these 4-point functions are scale invariant and satisfy equation \eqref{eq:conf4} and are thus both conformally invariant. This is not surprising as the resulting trispectrum is a function of the de Sitter invariant distances. 

The 4-point function of scalars of conformal dimension $\Delta$ takes the form
\be
F\left(\frac{r_{13} r_{24}}{r_{12} r_{34}}, \frac{r_{23} r_{41}}{r_{12} r_{34}} \right) \prod_{i<j} r_{ij}^{-2 \Delta/3} \;,
\ee
where $r_{ij}$ is the distance between point $i$ and point $j$.
It is a function of two variables, the harmonic ratios, instead of five variables as in the absence of conformal invariance.  Notice that with four points, one can indeed build ratios of the de Sitter invariant distances which are time-independent, and thus not fixed by the time evolution. The spatial dependence one gets
is, of course, the one of the harmonic ratios above.
Derivative operators will give a 4-point function which is suppressed when one of the points, say $1$ goes to infinity. In this limit the function $F$ becomes $F(r_{24}/r_{34}, r_{23}/r_{34})$, where the arguments are still generic, but they have to satisfy the triangle inequality. We conclude that the function $F$ can be large only when its arguments violate the triangle inequality. It would be interesting to study the most generic conformal 4-point function in Fourier space, as it is closer to data analysis.

Notice that, even in the presence of couplings with the inflaton, it is still possible for the correlation functions of the additional fields to enjoy the full $SO(4,1)$ symmetry. This happens if the action for perturbations preserves an approximate Lorentz symmetry, with the (inevitable) Lorentz breaking terms suppressed by an energy scale which is higher than for the invariant ones \cite{Senatore:2010wk}. 

The analysis of this paper has been restricted to inflation, but much of it can be applied to models in which the $SO(4,1)$ symmetry is not the isometry group of de Sitter, but the unbroken subgroup of the 4-dimensional conformal group $SO(4,2)$ \cite{Rubakov:2009np,Creminelli:2010ba,Hinterbichler:2011qk}; work is in progress in this direction. 

If experiments will detect primordial non-Gaussianity, our results will have clear implications. Correlation functions which are not conformal invariant cannot be generated by a sector which is decoupled from the inflaton. Therefore the inflaton itself (or possibly some other field with a large coupling with it) must be responsible for the observed scalar perturbations. In particular a 3-point function of ``equilateral" form, typical of inflationary models with small speed of sound (see for example \cite{Cheung:2007st}), is not conformal and its detection would unavoidably tell us that the inflaton plays a role is producing density perturbations. This nicely complements the general results one can prove for any single-field model, which constrain the behavior of the $n$-point functions of $\zeta$ in the squeezed limit \cite{astro-ph/0210603,astro-ph/0407059,arXiv:0709.0295}. These consistency relations imply that a detection of a sizable correlation function in the squeezed limit would unavoidably tell us that density perturbations cannot be generated solely by the inflaton. These two complementary arguments will allow to derive, independently of any specific models, basic information about the mechanism responsible for primordial perturbations.

\noindent
{\em Acknowledgements.}
It is a pleasure to thank Guido D'Amico, Edi Gava, Lam Hui, Juan Maldacena, Marcello Musso,  Alberto Nicolis, Jorge Nore\~na, Enrico Pajer, Aseem Paranjape, Slava Rychkov, Leonardo Senatore, Sergei Sibiryakov, Kendrik Smith, Enrico Trincherini, Filippo Vernizzi and especially Marko Simonovic and Matias Zaldarriaga for useful discussions. I thank the warm hospitality of the Institute for Advanced Studies in Princeton, where this work was initiated.



\begin{thebibliography}{99}

\bibitem{Creminelli:2010ba}
  P.~Creminelli, A.~Nicolis and E.~Trincherini,
  JCAP {\bf 1011}, 021 (2010)
  [arXiv:1007.0027 [hep-th]].
  
\bibitem{Maldacena:2011nz}
  J.~M.~Maldacena and G.~L.~Pimentel,
  arXiv:1104.2846 [hep-th].
  
\bibitem{Antoniadis:2011ib}
  I.~Antoniadis, P.~O.~Mazur and E.~Mottola,
  arXiv:1103.4164 [gr-qc].
  
\bibitem{Spradlin:2001pw}
  M.~Spradlin, A.~Strominger and A.~Volovich,
  ``Les Houches lectures on de Sitter space,''
  arXiv:hep-th/0110007.
  
\bibitem{hep-ph/0109214} 
  K.~Enqvist and M.~S.~Sloth,
  Nucl.\ Phys.\ B\ {\bf 626}, 395  (2002)
  [hep-ph/0109214].
  
\bibitem{Lyth:2001nq}
  D.~H.~Lyth and D.~Wands,
  Phys.\ Lett.\  B {\bf 524}, 5 (2002)
  [arXiv:hep-ph/0110002].
  
\bibitem{Dvali:2003em}
  G.~Dvali, A.~Gruzinov and M.~Zaldarriaga,
  Phys.\ Rev.\  D {\bf 69}, 023505 (2004)
  [arXiv:astro-ph/0303591].
      
\bibitem{Lyth:2004gb}
  D.~H.~Lyth, K.~A.~Malik and M.~Sasaki,
  JCAP {\bf 0505}, 004 (2005)
  [arXiv:astro-ph/0411220].
  
\bibitem{Zaldarriaga:2003my}
  M.~Zaldarriaga,
  Phys.\ Rev.\  D {\bf 69}, 043508 (2004)
  [arXiv:astro-ph/0306006].
  
\bibitem{Seery:2008qj}
  D.~Seery, K.~A.~Malik and D.~H.~Lyth,
  JCAP {\bf 0803}, 014 (2008)
  [arXiv:0802.0588 [astro-ph]].
  
\bibitem{Senatore:2010wk}
  L.~Senatore and M.~Zaldarriaga,
  arXiv:1009.2093 [hep-th].
  
\bibitem{Dalal:2007cu}
  N.~Dalal, O.~Dore, D.~Huterer and A.~Shirokov,
  Phys.\ Rev.\  D {\bf 77}, 123514 (2008)
  [arXiv:0710.4560 [astro-ph]].
  
\bibitem{Byrnes:2006vq}
  C.~T.~Byrnes, M.~Sasaki and D.~Wands,
  Phys.\ Rev.\  D {\bf 74}, 123519 (2006)
  [arXiv:astro-ph/0611075].
  
\bibitem{Rubakov:2009np}
  V.~A.~Rubakov,
  JCAP {\bf 0909}, 030 (2009)
  [arXiv:0906.3693 [hep-th]].
  
\bibitem{Hinterbichler:2011qk}
  K.~Hinterbichler and J.~Khoury,
  arXiv:1106.1428 [hep-th].
  
\bibitem{Cheung:2007st}
  C.~Cheung, P.~Creminelli, A.~L.~Fitzpatrick, J.~Kaplan and L.~Senatore,
  JHEP {\bf 0803}, 014 (2008)
  [arXiv:0709.0293 [hep-th]].
  
\bibitem{astro-ph/0210603} 
  J.~M.~Maldacena,
  JHEP\ {\bf 0305}, 013  (2003)
  [astro-ph/0210603].
  
\bibitem{astro-ph/0407059} 
  P.~Creminelli and M.~Zaldarriaga,
  JCAP\ {\bf 0410}, 006  (2004)
  [astro-ph/0407059].
  
\bibitem{arXiv:0709.0295} 
  C.~Cheung, A.~L.~Fitzpatrick, J.~Kaplan and L.~Senatore,
  JCAP\ {\bf 0802}, 021  (2008)
  [arXiv:0709.0295 [hep-th]].
 



\end{thebibliography}
\end{document}